\documentclass[twocolumn]{aastex631}
\usepackage{upgreek}

\begin{document}

\title{First Detections of PN, PO, and PO$^{+}$ toward a Shocked Low-mass Starless Core}

\author[0000-0002-9485-4394]{Samantha Scibelli}\email{sscibell@nrao.edu}
              \thanks{Jansky Fellow of the National Radio Astronomy Observatory.}
\affiliation{National Radio Astronomy Observatory, 520 Edgemont Road, Charlottesville, VA 22903, USA}

\author[0000-0002-6389-7172]{Andr\'es Meg\'ias}
\affiliation{Centro de Astrobiolog\'ia (CAB), CSIC-INTA, Carretera de Ajalvir, km 4, E-28805 Torrej\'on de Ardoz, Spain}

\author[0000-0003-4493-8714]{Izaskun Jim\'enez-Serra}
\affiliation{Centro de Astrobiolog\'ia (CAB), CSIC-INTA, Carretera de Ajalvir, km 4, E-28805 Torrej\'on de Ardoz, Spain}

\author[0000-0002-0133-8973]{Yancy Shirley}
\affiliation{Steward Observatory,  University of Arizona, 933 North Cherry Avenue,
Tucson, AZ 85721, USA}

\author[0000-0002-8716-0482]{Jennifer Bergner}
\affiliation{Department of Chemistry, University of California, Berkeley, California 94720-1460, USA}

\author[0000-0002-1095-9368]{Judit Ferrer Asensio}
\affiliation{RIKEN Cluster for Pioneering Research, Wako-shi, Saitama, 351-0106, Japan}

\author[0000-0001-7723-8955]{Robin T. Garrod}
\affiliation{Departments of Astronomy \& Chemistry, University of Virginia, Charlottesville, VA 22904, USA}   

\author[0000-0001-6551-6444]{M\'elisse Bonfand}
\affiliation{Departments of Astronomy \& Chemistry, University of Virginia, Charlottesville, VA 22904, USA}    

\author[0009-0002-1883-150X]{Anissa Pokorny-Yadav}
\affiliation{Department of Astronomy, University of California, Berkeley, California 94720-1460, United States, USA}
        


\begin{abstract}

Phosphorus is a key element that plays an essential role in biological processes important for living organisms on Earth.
The origin and connection of phosphorus-bearing molecules to early Solar system objects and star-forming molecular clouds is therefore of great interest, yet there are limited observations throughout different stages of low-mass ($M < $ a few M$_\odot$) star formation. Observations from the Yebes 40 m and IRAM 30 m telescopes detect for the first time in the 7mm, 3mm, and 2mm bands multiple transitions of PN and PO, as well as a single transition of PO$^{+}$, toward a low-mass starless core. 
The presence of PN, PO and PO$^{+}$ is kinematically correlated with bright SiO(1-0) emission. Our results reveal not only that shocks are the main driver of releasing phosphorus from dust grains and into the gas-phase, but that the emission originates from gas not affiliated with the shock itself, but quiescent gas that has been shocked in the recent past. 
From radiative transfer calculations, the PO/PN abundance ratio is found to be $3.1^{+0.4}_{-0.6}$, consistent with other high-mass and low-mass star-forming regions. This first detection of PO$^{+}$ toward any low-mass star-forming region reveals a PO$^{+}$/PO ratio of $0.0115^{+0.0008}_{-0.0009}$, a factor of ten lowerk than previously determined from observations of a Galactic Center molecular cloud, suggesting its formation can occur under more standard Galactic cosmic-ray ionization rates. These results motivate the need for additional observations that can better disentangle the physical mechanisms and chemical drivers of this precursor of prebiotic chemistry. 
\end{abstract}




\section{Introduction} \label{sec:intro}

A necessary element for life on Earth, phosphorus is a critical component in DNA, RNA, ATP, and a host of other biological molecules (e.g., \citealt{2005Chem...Macia, 2006Phis...Shwartz}). 
The best constraints on how phosphorus was distributed within the Interstellar Medium (ISM) and during the formation of our own solar system, as well as other Sun-like stars and planetary systems, are made from observations of P-bearing molecules, such as phosphorus mononitride, PN, phosphorus monoxide, PO, and the phosphorus oxide ion, PO$^{+}$ (see recent review by \citealt{2024FrASS..1151127F}). 

Although PN has been known for decades to be present in shells of evolved stars \citep{2000A&AS..142..181C, 2007ApJ...666L..29T} as well as massive star-forming regions \citep{1987ApJ...321L..75T, 1987ApJ...321L..81Z, 1990ApJ...365..569T}, there are limited constraints for PN (as well as PO and PO$^{+}$) in low-mass ($M < $ a few M$_\odot$) star-forming environments, with only a handful of detections toward low-mass protostars and outflows \citep{2011PASJ...63L..37Y, 2016MNRAS.462.3937L, 2019ApJ...884L..36B, 2022ApJ...934..153W}. We do know, however, that PN is present in quiescent high-mass dense cores \citep{2016ApJ...822L..30F}, as well as the central star-forming region of the distant starburst galaxy NGC 253\,\citep{2022A&A...659A.158H}. Still, there are no published constraints for PN, PO or the PO$^{+}$ ion in the earlier, nearby ($<400$\,pc), cold ($\sim$\,10\,K), and moderately dense ($n \sim 10^{5}$\,cm$^{-3}$) low-mass starless and prestellar core stages. 

Observations of gas-phase P-bearing molecules are very well correlated both kinematically and spatially with the shock tracer SiO, supporting the idea that phosphorus is first released via grain sputtering in shocked regions\,\citep{2016MNRAS.462.3937L, 2018MNRAS.475L..30R, 2018MNRAS.476L..39M, 2019MNRAS.489.4530F, 2022ApJ...927....7B, 2024A&A...687A..75L}. While it is still not certain what the main carrier of phosphorus is in dust grains, once P is released it can form into molecules via ion-neutral or neutral-neutral gas-phase reactions (e.g., \citealt{2018ApJ...862..128J, 2021ApJ...922..169G, 2024ApJ...963..142G}).

At a distance of 294\,pc (see \citealt{2021A&A...645A..55P}), the low-mass star-forming NGC\,1333 region within the Perseus Molecular Cloud is known to host quiescent starless and prestellar cores as well as active protostars, which are associated with dynamic CO outflows and extended SiO emission red{\citep{1998ApJ...504L.109L, 1999A&A...343..585C, 2000A&A...361..671K, 2013ApJ...774...22P}. The widespread ($\sim0.4$\,pc across) and abundant SiO emission mapped by \cite{1998ApJ...504L.109L} shows a narrow line ($\Delta v = 0.4-1.2$\,km/s) `ambient component', separate from additional broad-line features associated with the high-velocity protostellar outflows, believed to be tracing rather the interaction of these outflows in the region with the dense gas. \cite{1999A&A...343..585C} instead attributed the narrow-line SiO to fossil shocks from the protostellar outflows, or slower shocks that would require a significant amount of silicon on dust grain mantles. More recently, \cite{2022MNRAS.512.5214D} hypothesized that this widespread narrow SiO is from expanding bubbles, or `large-scale turbulent cells' \citep{2019ApJ...876..108D}, that collided with the cloud and triggered successive shocks. The origin of these bubbles, however, is less clear.

Directly coincident with this ambient SiO emission in  NGC\,1333 is a `prestellar' core $\sim$6200 au or 21$^{''}$ in size, labeled as such and indexed `326' in the \textit{Herschel} Perseus core catalog of \cite{2021A&A...645A..55P} based firstly on the fact that no internal source of energy (e.g., protostar) is present and second, on a virial analysis that found the core's self-gravity exceeds its pressure support. We stress, however, that because the dynamical properties and other evolutionary parameters (e.g., infall signatures and a CO depletion factor) are not known, we will continue to refer to `Pers326' as starless, not prestellar. The coordinates of Pers326 also coincides with an ammonia (NH$_3$) peak \citep{2008ApJS..175..509R, 2019ApJ...874..147K} and thus provides additional information such as the core's kinetic temperature ($T_\mathrm{k}\sim 12$\,K). 

This criterion outlined above (\textit{Herschel} plus NH$_3$ observations) led to the investigation of the complex chemical inventory of Pers326. It was included in a sample of 35 total starless and prestellar cores in the Perseus Molecular Cloud surveyed with the Arizona Radio Observatory (ARO) 12\,m dish, presenting high abundances of the simple complex organic molecules (COMs) CH$_3$OH and CH$_3$CHO \citep{2024MNRAS.533.4104S}. Follow-up observations were done with the Yebes Observatory 40\,m telescope's Q-band receiver to search for higher complexity COMs in a sub-sample of the 15 cores with bright CH$_3$OH and CH$_3$CHO, including Pers326 (also presented in \citealt{2024MNRAS.533.4104S}). The wide-band nature of the Q-band observations led to the serendipitous detection of the PN(1-0) line, concurrent with bright SiO(1-0) emission, which we present here and confirm with additional observations at higher frequencies. 

In section\,\ref{sec:obs} we describe our observations, in section\,\ref{sec:detections} we present our detections, in section\,\ref{sec:columndensities} we calculate column densities, and in section\,\ref{sec:discuss} we put our observations in context with literature values. Lastly, in section\,\ref{sec:conclude} we conclude and discuss implications for future work.

\section{Observations} \label{sec:obs}

We present single-pointing observations of the starless core Pers326 (RA: $03^\mathrm{h}29^\mathrm{m}08.97^\mathrm{s}$, Dec: $+31^\mathrm{o}15{\arcmin}17.2\arcsec$) from both the 40\,m radio telescope of the Yebes Observatory (Yebes 40\,m), in Guadalajara, Spain, as well as the 30\,m radio telescope of the Institut de Radioastronomie Millimétrique (IRAM 30\,m) in Pico Veleta, Spain. 

The Yebes 40\,m observations were taken during the spring 2022 and spring 2023 seasons (projects 22A022 and 23A025; PI: Scibelli) with the dual (horizontal and vertical) linear polarization Q-band receiver \citep{2021A&A...645A..37T} using the frequency switching technique with a standard throw of 10.52 MHz. The wide-band nature of the receiver allows for a total bandwidth of 18.5\,GHz spanning from 31.5 to 50\,GHz (6 to 9\,mm) with a resolution of 38.0\,kHz (0.38 to 0.23\,km/s).
These data were originally presented with more detail in \cite{2024MNRAS.533.4104S}. 

Follow-up IRAM 30\,m observations were taken in October and December 2024 (project 024-24; PI: Scibelli) with the broad-band EMIR 2\,mm and 3\,mm receivers \citep{2012A&A...538A..89C}. The FTS50 backend provided a spectral resolution of 50\,kHz, corresponding to channels of 0.1 to $0.2$\,km/s. The position-switching technique was chosen for these observations (with an `off' position of RA: $03^\mathrm{h}28^\mathrm{m}00^\mathrm{s}$, Dec: $+31^\mathrm{o}19{\arcmin}07\arcsec$, the same as used in ARO 12\,m observations from \citealt{2024MNRAS.533.4104S}) in order to achieve more stable baselines. Pointing and focusing was done on Mars and/or Jupiter every one to two hours depending on the weather. Additionally, the nearby protostar IRAS\,4A (see Figure\,\ref{fig:outflow_info}) was chosen as a line calibrator, since it has been extensively observed with the IRAM 30\,m for the ASAI large program \citep{2018MNRAS.477.4792L}. We find conditions stable, with antenna temperatures ($T_\mathrm{A}^*$) varying by less than 10\%.  

For both the Yebes 40\,m and IRAM 30\,m datasets the reduction was done with the publicly available Python-based scripts\footnote{\url{https://github.com/andresmegias/gildas-class-python/}} developed by \cite{2023MNRAS.519.1601M}, which invokes the CLASS program of the GILDAS package (\citealt{2005sf2a.conf..721P, 2013ascl.soft05010G}) for the combining of the spectra (again, see \citealt{2024MNRAS.533.4104S} for more detail). Data were inspected, baselined, combined, and put on the main beam temperature, $T_\mathrm{mb}$, scale based on the public efficiency measurements given for each respective telescope and receiver. Gaussian fitting of spectral lines was done with the Python `Pyspeckit' package \citep{2011ascl.soft09001G, 2022AJ....163..291G}. The lines presented and analyzed here are listed in Table\,\ref{table:1} and plotted in Figures\,\ref{fig:yebesSiO},\,\ref{fig:yebesPN},\,\ref{fig:PO} and \ref{fig:POplus}.

\begin{figure}
\begin{center}$
\centering
\begin{array}{c}
\includegraphics[width=80mm]{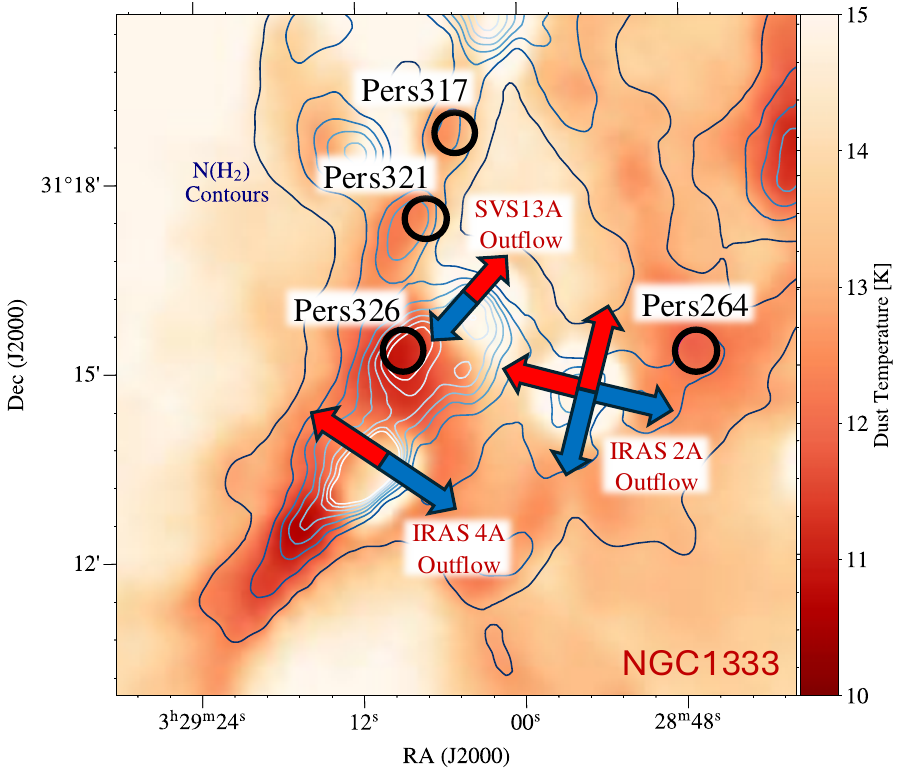} 
\end{array}$
\end{center}
    \caption{ Dust temperature (color scale) and $N$(H$_2$) column density (contours: $1, 2, 3, 4, 5, 6, 7, \mathrm{and}\, 8 \times10^{22}$\,cm$^{-2}$) \textit{Herschel} maps of NGC\,1333 region in Perseus (\citealt{2021A&A...645A..55P}). Included are the locations of four starless and prestellar cores, Pers264, Pers317, Pers321 and Pers326 (black circles 38$''$ in size) and the locations of the prominent CO outflows in the region, SVS\,13A, IRAS\,4A, and IRAS\,2A, as described in \citealt{2013ApJ...774...22P} (note: the length of jet arrows are not to scale).} \label{fig:outflow_info}
\end{figure}

\begin{table*}
\caption{Molecular lines analyzed}            
\label{table:1}     
\setlength{\tabcolsep}{5pt}
\begin{center}                     
\begin{tabular}{c c c c c c c c c c  }       
\hline\hline                 
 & & & & & & \multicolumn{4}{c }{Source: Pers326}  \\
 
 Species & Transition & $\nu$ & $E_\mathrm{u}/k$ & $g_\mathrm{u}$ & $A_\mathrm{ul}$ & \textit{I} & FWHM & Vel & rms\\  

     & & [GHz]& [K] &   & [s$^{-1}$] & [mK km s$^{-1}$] & [km s$^{-1}$] & [km s$^{-1}$] & [mK] \\ 
\hline                        

 \multicolumn{10}{c}{\textbf{Yebes 40\,m}} \\ 
 \hline
 
 SiO & 1-0 & 43.4238550(30)  & 2.1 & 3 & 3.0$\times 10^{-6}$  & 1330.938(1) & 0.900(1) & 7.6506(2) & 3.6 \\ 
 PN  & $J$ = 1-0, $F$ = 1-1 & 46.9889812(8)  & 2.3  & 3  & 1.0$\times 10^{-6}$  & -- & --& --& 3.0  \\
   & $J$ = 1-0, $F$ = 2-1 & 46.9905445(7)  & 2.3 & 5 & 1.0$\times 10^{-6}$  &  33.7(2) & 1.36(10) & 7.80(5) & 3.0 \\   
   & $J$ = 1-0, $F$ = 0-1 & 46.9928270(12)  & 2.3 & 1 & 1.0$\times 10^{-6}$  & -- & --& --&  3.0  \\   
PO$^+$ & 1-0 & 47.0242519(69) & 2.3 & 3 & 4.8$\times 10^{-6}$ & -- & -- & -- & 3.0\\
\hline 
 \multicolumn{10}{c}{\textbf{IRAM 30\,m}} \\ 
 \hline
  PN & $J$ = 2-1, $F$ = 2-2 & 93.9782063(15)   &  6.8 & 5 & 2.4$\times 10^{-6}$  & -- & --& --& 4.1\\  
   & $J$ = 2-1, $F$ = 1-0 & 93.9784738(12)  &  6.8 & 3 & 5.4$\times 10^{-6}$   & -- & --& --& 4.1 \\ 
   & $J$ = 2-1, $F$ = 2-1 & 93.9797695(11)  & 6.8 & 5 & 7.3$\times 10^{-6}$  & 48.9(2) & 0.99(2) & 7.37(1) & 4.1\\ 
   & $J$ = 2-1, $F$ = 3-2 & 93.9798901(12) &  6.8 & 7 &  9.7$\times 10^{-6}$  & -- & --& --& 4.1 \\ 
  & $J$ = 2-1, $F$ = 1-2 & 93.9807563(23)  & 6.8 & 3 & 2.7$\times 10^{-7}$  & -- & --& --& 4.1 \\ 
   & $J$ = 2-1, $F$ = 1-1 & 93.9823196(15) & 6.8 & 3 & 4.0$\times 10^{-6}$  & -- & --& --& 4.1\\ 
   
  & $J$ = 3-2, $F$ = 3-3 & 140.9660092(22) & 13.5 & 7& 3.9$\times 10^{-6}$  & -- & --& --& 6.2 \\
 & $J$ = 3-2, $F$ = 2-1 & 140.9674256(17)& 13.5 & 5 & 3.0$\times 10^{-5}$  & 32.00(2) & 1.08(2) & 7.03(1) & 6.2 \\
 & $J$ = 3-2, $F$ = 3-2 & 140.9676931(16)& 13.5& 7& 3.1$\times 10^{-5}$  & -- & --& --& 6.2\\
 & $J$ = 3-2, $F$ = 4-3 & 140.9677647(17) & 13.5 & 9 & 3.5$\times 10^{-5}$  & -- & --& --& 6.2\\
  & $J$ = 3-2, $F$ = 2-2 & 140.9699756(17)& 13.5 & 5 & 5.5$\times 10^{-6}$  & -- & --& --& 6.2 \\
  
 PO & $J$ = 5/2-3/2, $\Omega$ = 1/2, &    &  &   &  &   &  &  &  \\ 
  & $F$ = 2-2, $l=f$ & 108.707192(20)  & 8.4 & 5 & 2.1$\times 10^{-6}$  & -- & --& --& 7.8\\  
 &   $F$ = 3-2, $l=e$ & 108.998445(20)   & 8.4 & 7 & 2.1$\times 10^{-5}$   & 28.98(2) & 0.58(1) & 7.621(4) & 6.8 \\ 
 &  $F$ = 2-1, $l=e$ & 109.045396(20)   & 8.4 & 5 & 1.9$\times 10^{-5}$  & 18.34(2) & 0.49(1) & 7.601(5) & 6.6 \\ 
  &  $F$ = 3-2, $l=f$ & 109.206200(20)  & 8.4 & 7 &2.1$\times 10^{-5}$  & 28.56(1) & 0.52(1) & 7.611(4) & 7.7 \\ 
  &  $F$ = 2-2, $l=e$ & 109.271376(20)   & 8.4 & 5 & 2.1$\times 10^{-6}$ & -- & --& --& 8.4 \\ 
  &  $F$ = 2-1, $l=f$ & 109.281189(20)   & 8.4 &  5 &1.9$\times 10^{-5}$  & 20.01(2) & 0.53(1) & 7.62(1) & 8.5 \\ 
  
  & $J$ = 7/2-5/2, $\Omega$ = 1/2,  &   &   &  &   &   &  & &  \\
    &$F$ = 3-3, $l=f$ & 152.389058(20) & 15.7 & 7& 3.0$\times 10^{-6}$  & -- & --& --& 6.8\\
  &  $F$ = 4-3, $l=e$ & 152.656979(20) & 15.7 & 9& 6.3$\times 10^{-5}$  & 13.82(2) & 0.43(1) & 7.635(5) & 6.0 \\
  & $F$ = 3-2, $l=e$ & 152.680282(20) & 15.7 &7 & 6.0$\times 10^{-5}$  & -- & --& --& 6.9 \\
  &  $F$ = 4-3, $l=f$ & 152.855454(20) & 15.8 &9 & 6.3$\times 10^{-5}$  & -- & --& --& 7.2 \\
  &  $F$ = 3-2, $l=f$ & 152.888128(20) & 15.7& 7& 6.0$\times 10^{-5}$  & 10.46(2) & 0.43(1) & 7.607(6) & 8.4 \\
  &  $F$ = 3-3, $l=e$ & 152.953247(20) & 15.7 & 7& 3.0$\times 10^{-6}$  & -- & --& --& 6.1 \\
  PO$^+$ & 2-1 & 94.0477984(132) & 6.8 & 5 & 4.6$\times 10^{-5}$ & 6.20(5) & 0.39(3) & 7.64(1) & 3.0\\
  & 3-2 & 141.0699343(183) & 13.5 & 7 & 1.7$\times 10^{-4}$ & -- & -- & -- & 5.0\\
\hline                                   
\end{tabular}
\end{center}
\tablecomments{ Line parameters include frequency, $\nu$, upper energy, $E_\mathrm{u}/k$, upper state degeneracy, $g_u$, and spontaneous emission coefficient, $A_\mathrm{ul}$, the majority of which are referenced from the Cologne Database for Molecular Spectroscopy (CDMS) catalog (\url{https://cdms.astro.uni-koeln.de};  \citealt{2001A&A...370L..49M, 2005JMoSt.742..215M, 2016JMoSp.327...95E}), except SiO which comes from SLAIM entry in SPLATALOGUE (\citealt{2007AAS...21113211R}, \url{https://splatalogue.online}) and PO$^+$ which comes from the JPL line catalog \citep{1998JQSRT..60..883P} . Specific laboratory data is detailed in the following work: \cite{1968ZNatA..23..777T, HoeftTiemannTörring+1972+703+704,1983JChPh..79..629K, 1991JChPh..94.3504P, 2006JMoSt.780..260C, 2013JPCA..11713843M}. The Gaussian fitting parameters for source Pers326 are also listed next to each line transition, where `\textit{I}' represents the integrated intensity on the main temperature beam scale, `FWHM' is the full width at half maximum linewidth, `Vel' is the line-of-sight centroid velocity, and `rms' is the 1$\sigma$ standard deviation in the noise level. Numbers in parentheses denote 1$\sigma$ uncertainties in unit of the last quoted digit.  
}
\end{table*}

\section{Detections and Line Properties} \label{sec:detections}

The wide-bandwidth in the Yebes 40\,m data led to the serendipitous detection of the PN(1-0) line at 46.990\,GHz toward the Pers326 starless core. A total of 15 cores bright in CH$_3$OH and CH$_3$CHO were observed in this sample \citep{2024MNRAS.533.4104S}, and so PN was also searched for in these cores with no success (see Table\,\ref{table:2}). Pers326 is the only core with strong ($T_\mathrm{mb}$ $>$ 1\,K) SiO(1-0) emission at 43.423855\,GHz\footnote{Note that if one were to use the measured frequency of 43.42376\,GHz from the JPL line list within the SPLATALOGUE database, SiO would be blue-shifted from the PN and PO emission by $\sim$0.8\,km/s. This JPL entry has a roughly order of magnitude larger error than the entry in SLAIM, i.e., 50\,kHz versus 3\,kHz, respectively. One should thus be careful when extracting rest frequencies from SPLATALOGUE, and use those entries with smaller associated errors. 
 }\,(Figure\,\ref{fig:yebesSiO}). This emission is only slightly blue-shifted, to 7.6\,km/s, from the PN(1-0) $v_\mathrm{lsr}$ of 7.8\,km/s, but given the coarse resolution of Yebes at these frequencies ($\sim$0.2\,km/s) we conclude they are consistent (Table\,\ref{table:1}).
A few other starless and prestellar cores reside nearby spatially to Pers326 with SiO detections, Pers264, Pers317 and Pers321 (see Figure\,\ref{fig:outflow_info}), but the emission is much weaker ($T_\mathrm{mb}$ $<$ 0.3\,K) and offset in $v_\mathrm{lsr}$ (Figure\,\ref{fig:yebesSiO}). For Pers264 the SiO(1-0) is blue-shifted more dramatically, to $\sim$2\,km/s, suggesting this emission is not associated with the core itself (at a similar $v_\mathrm{lsr}$ to Pers326) but with the larger-scale surrounding SiO gas (see Figure\,\ref{fig:yebesSiO}). Given the strong SiO(1-0) and PN(1-0) correlation in Pers326, we suggest the origin of this gas-phase PN is related to shock chemistry (see section\,\ref{sec:discuss} for more discussion). 

Only one transition of PN was available in Q-band, therefore dedicated follow-up observations with the IRAM\,30\,m allowed us to not only confirm the detection of PN by observing two other higher energy transitions ($J$\,=\,2-1 and $J$\,=\,3-2), but they also allowed for the firm detection of multiple transitions of the chemically related P-bearing molecule PO (Figure\,\ref{fig:PO}). We can pick out above the noise levels ($>3\sigma$) individual hyperfine structure for six PO lines (Table\,\ref{table:2}), and we tentatively detect another two transitions (at 152.680\,GHz and 152.855\,GHz) out of the total twelve observed (see Figure\,\ref{fig:PO}). The detected and tentatively detected lines are those with the faster (i.e., larger) $A_\mathrm{ul}$ values at $>1\times10^{-6}$\,s$^{-1}$, compared to the non-detections with $A_\mathrm{ul}$\,$<1\times10^{-6}$\,s$^{-1}$, in agreement with predicted line intensities (more discussion in section\,\ref{sec:columndensities}).

\begin{figure}[tbh]
\begin{center}$
\centering
\begin{array}{c}
\includegraphics[width=75mm]{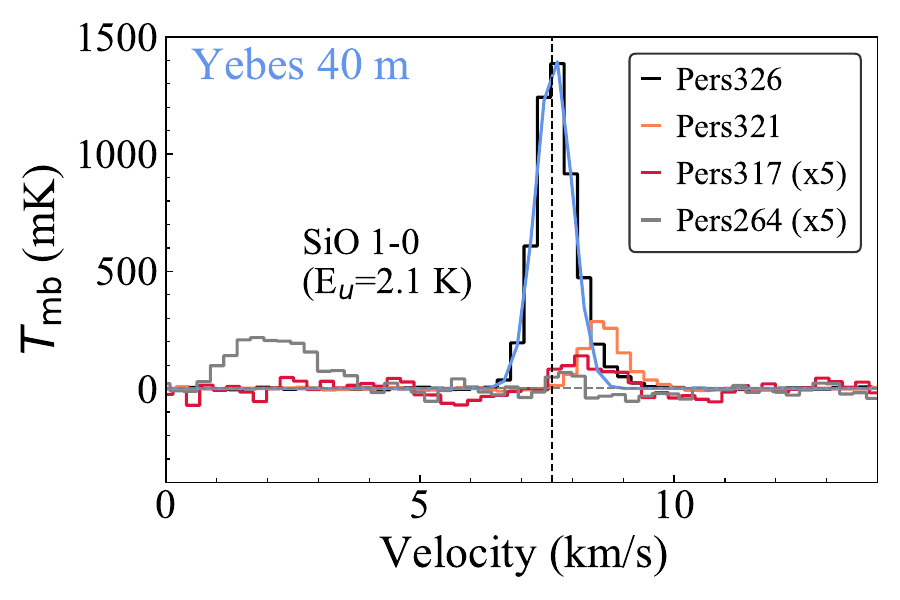}  
\end{array}$
\end{center}
    \caption{SiO(1-0) emission observed toward Pers326 with the Yebes 40\,m telescope (black) and a Gaussian fit to the line (blue). The vertical dashed lines correspond to the $v_\mathrm{lsr}$ of 7.6\,km/s. Additional SiO emission toward the other nearby starless cores Pers321 (orange), Pers317 (red) and Pers264 (gray) are also plotted for reference. 
    }
    \label{fig:yebesSiO}
\end{figure}

\begin{figure}[tbh]
\begin{center}$
\centering
\begin{array}{c}
\includegraphics[width=75mm]{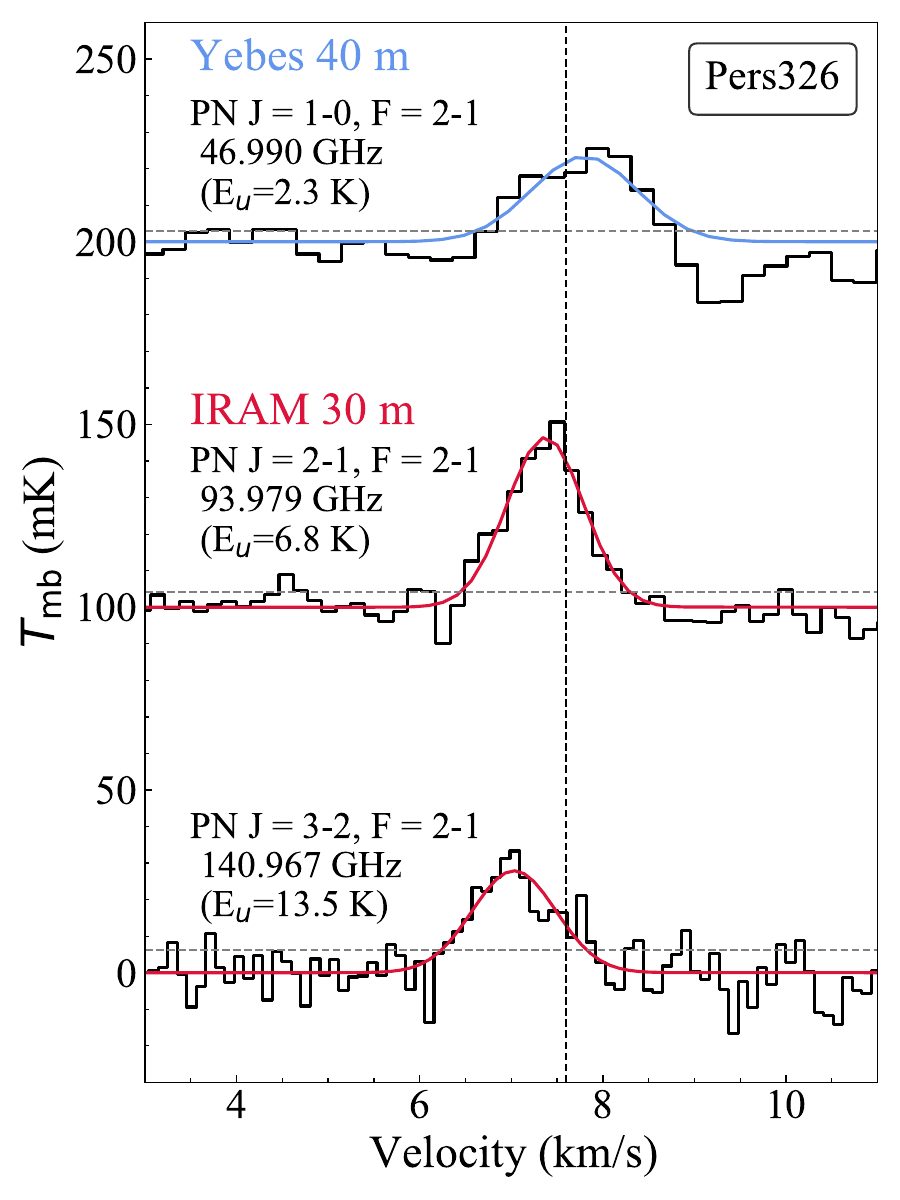} 
\end{array}$
\end{center}
    \caption{PN emission centered on the $F$ = 2-1 hyperfine component for each of the $J$ transitions observed toward Pers326 with both the Yebes 40\,m and IRAM 30\,m. As solid and blue and red curves are the Gaussian fits to the lines, the vertical dashed line shows a $v_\mathrm{lsr}$ of 7.6 km/s, and the horizontal gray line shows $1\sigma$ noise level. Spectra are offset by 100\,mK for easier viewing. Annotated on the spectra are also the locations of the weaker and blended hyperfine components.} \label{fig:yebesPN}
\end{figure}

The $v_\mathrm{lsr}$ of all the detected PO transitions are remarkably consistent with each other and the SiO 1-0 emission, at 7.6\,km/s. The PO linewidths (FWHMs) range between $0.43 - 0.58$\,km/s, narrower than both the SiO and PN emission, with FWHM $\sim 1$\,km/s. Hyperfine blending in PN likely contributes to the offset in the $v_\mathrm{lsr}$ for the $J$ ladder $F$ = 2-1 transitions (see Figure\,\ref{fig:yebesPN}), but may also be due to uncertainty in the rest frequencies and/or multiple velocity components, which appear compact within the large single-dish beam of the Yebes 40\,m (38$''$). Follow-up observations at even higher spatial and spectral resolution would help to disentangle the cause of these offsets. 

We also search for and successfully identify the ion PO$^{+}$. The 1-0 transition lies in the Yebes 40m bandpass, whereas the 2-1 and 3-2 transition lies within our IRAM 30m bandpass (Table\,\ref{table:1}). We only detect the 2-1 line at 5$\sigma$ confidence, where $T_\mathrm{mb} = 15$\,mK at an rms level of 3.0\,mK (see Figure\,\ref{fig:POplus}). Assuming the same excitation conditions as PO (more discussion in section\,\ref{sec:columndensities}), we predict the 1-0 and 3-2 line intensities to be weaker, below the $3\sigma$ detection limit and consistent with our observations (Figure\,\ref{fig:POplus}). The $v_\mathrm{lsr}$ of the PO$^{+}$ 2-1 transition is in good agreement with the PO lines, with a slightly narrower linewidth, at 0.39\,km/s. It's worth noting that the Yebes resolution at the 1-0 line is coarser, at 0.25\,km/s (compared to the 2-1 line at 0.16\,km/s), and thus this line may also be resolved out.

\begin{figure} 
\begin{center}$
\centering
\begin{array}{c}
\includegraphics[width=75mm]{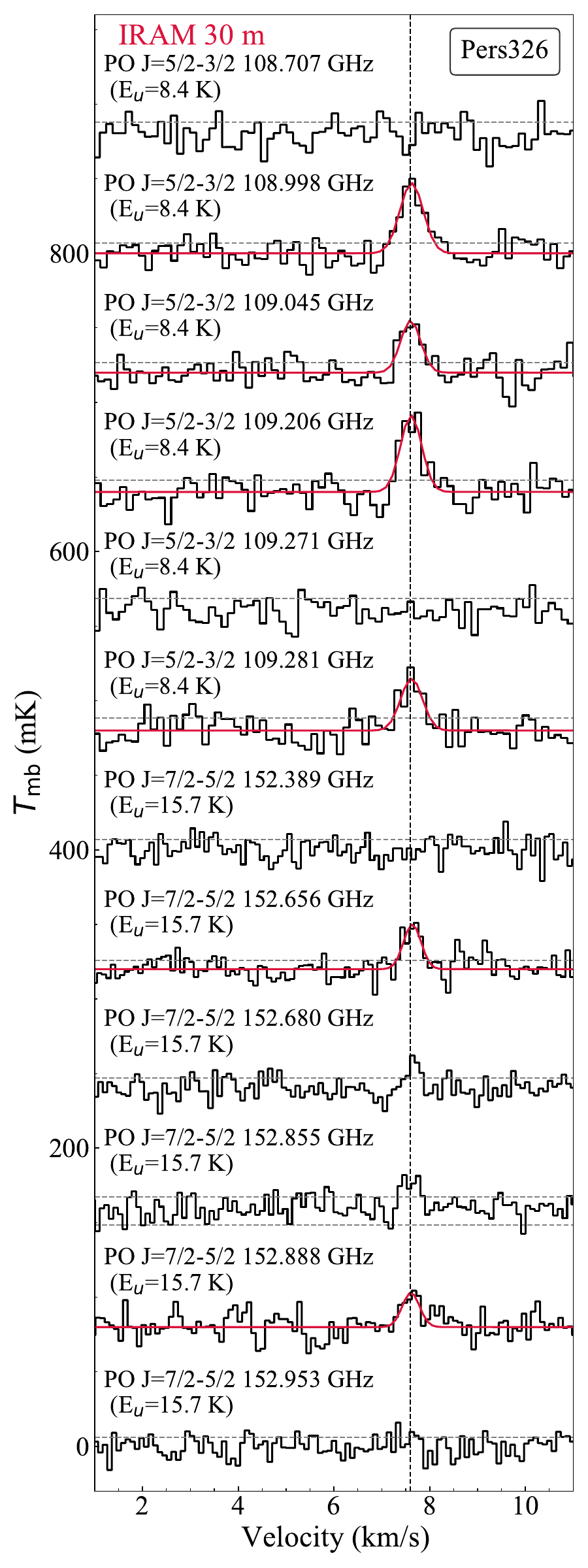} 
\end{array}$
\end{center}
    \caption{PO emission observed toward Pers326 with the IRAM 30\,m using both the 3\,mm and 2\,mm receivers. In solid red are the Gaussian fits to the detected ($>3\sigma$) lines and in dotted red are LTE fits at $T_\mathrm{ex}$ = 4.6\,K. The vertical dashed line shows the consistent $v_\mathrm{lsr}$ of 7.6\,km/s, and the horizontal gray line shows $1\sigma$ noise level. Spectra are offset by 80\,mK for easier viewing. Full quantum numbers for each transition are given in Table\,\ref{table:1}.} \label{fig:PO}
\end{figure}

\begin{figure}
\begin{center}$
\centering
\begin{array}{c}
\includegraphics[width=75mm]{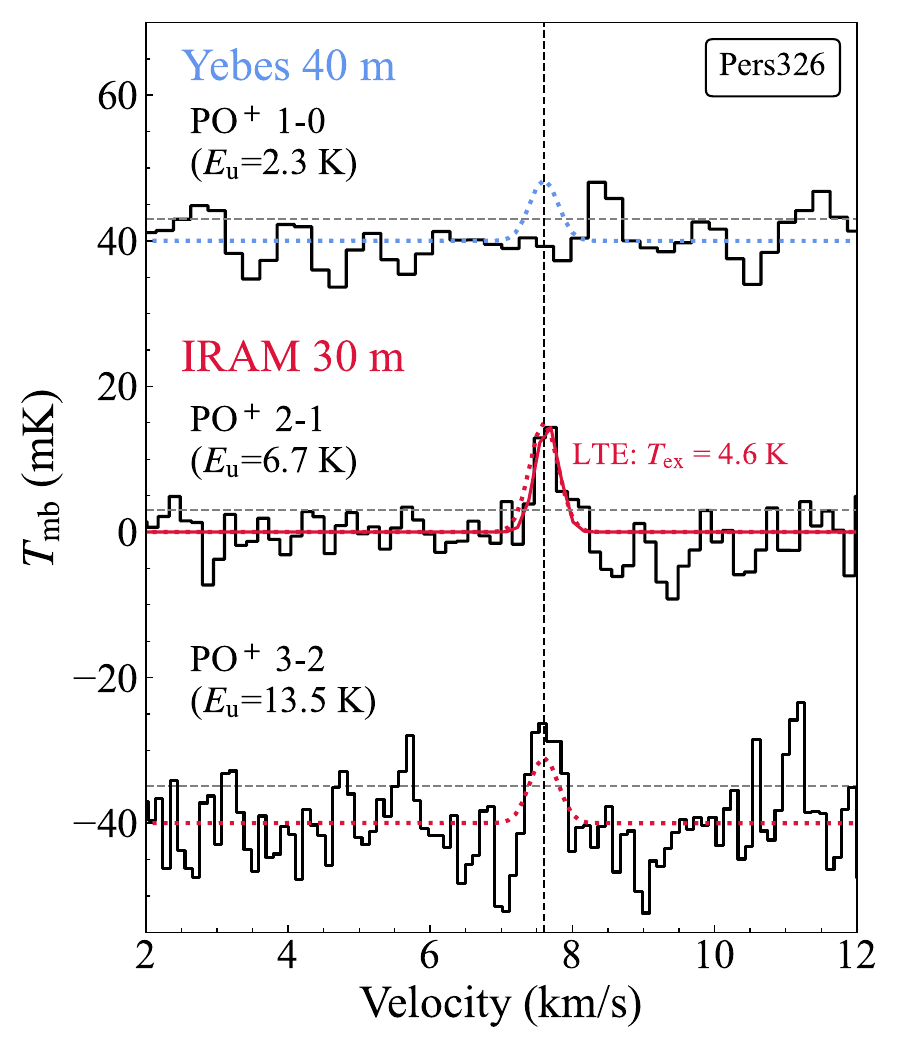} 
\end{array}$
\end{center}
    \caption{PO$^{+}$ transitions observed toward Pers326. In solid red is the Gaussian fit to the detected ($>3\sigma$) line and the dotted lines are LTE fits at $T_\mathrm{ex}$ = 4.6\,K. The vertical dashed line shows the consistent $v_\mathrm{lsr}$ of 7.6\,km/s, and the horizontal gray line shows $1\sigma$ noise level. Spectra are offset by 40\,mK for easier viewing.} \label{fig:POplus}
\end{figure}

\section{Column Densities} \label{sec:columndensities}

We use the rotation diagram method to constrain the PN and PO column densities by assuming we are in the optically thin limit and also in local thermodynamical equilibrium (LTE). The upper state column density can be calculated by: \begin{equation}
    {N}_\mathrm{u} = \frac{I}{h A_\mathrm{ul} f} \frac{u_\nu({T}_\mathrm{ex})}{[J_\nu({T}_\mathrm{ex}) - J_\nu({T}_\mathrm{cmb})]},
\end{equation} where $h$ is the Planck constant, $f$ is our (frequency-dependent) filling factor, $I$ is the integrated intensity of the line, $A_\mathrm{ul}$ is the spontaneous emission coefficient (or `Einstein A'), and $T_\mathrm{cmb}$ is the background temperature of 2.73\,K. We define both the Planck energy density, $u_\nu$, and the Planck function in temperature units, $J_\nu$, where $\nu$ is our line frequency, as: \begin{equation}
    u_\nu \equiv \frac{8 \uppi h \nu^3 }{c^3} \frac{1}{\exp{(h\nu/kT) - 1}}, 
\end{equation} \begin{equation}
    J_\nu \equiv \frac{h \nu }{k} \frac{1}{\exp{(h\nu/kT) - 1}}.
\end{equation} Our constants are, $c$, the speed of light, and $k$, the Boltzmann constant. For a total column density, $N_\mathrm{{tot}}$, \begin{equation} \label{eq:N}
   \frac{{N}_\mathrm{u}}{g_\mathrm{u}} = \frac{N_\mathrm{tot}}{{Q(T_\mathrm{ex})}} \exp{(-E_\mathrm{u} / k {T_\mathrm{ex}})},
\end{equation} where $g_\mathrm{u}$ is the upper state degeneracy, $E_\mathrm{u}$ is the upper state energy, and $Q(T_\mathrm{{ex})}$ is the partition function dependent on the excitation temperature of the molecule (that we calculate explicitly for the best-fit $T_\mathrm{{ex}}$). The natural logarithm of the left side of equation\,\ref{eq:N}, ln(${N}_\mathrm{u}/g_\mathrm{u}$), is plotted versus $E_\mathrm{u}$ so that the excitation temperature, ${T_\mathrm{ex}}$, is the inverse of the slope of the linear fit and the y-intercept finds ${N_\mathrm{tot}}$ \citep{1999ApJ...517..209G}. 

Figure\,\ref{fig:rot} shows, firstly, a least-squares fit to the PN data where $T_\mathrm{ex} = 7.6\pm$0.7\,K and 
$N_\mathrm{tot} = 0.84\pm0.12 \times 10^{12}$\,cm$^{-2}$.
A separate fit to the PO data finds a steeper slope where $T_\mathrm{ex} = 4.63\pm$0.13\,K and 
$N_\mathrm{tot} = 2.63\pm0.16 \times 10^{12}$\,cm$^{-2}$. Uncertainties in the fits were calculated based on Monte Carlo simulations sampling the data points as Gaussians and then analyzing the resulting distributions of the parameters using the richvalues Python package\footnote{https://github.com/andresmegias/richvalues}. While the spread in $E_\mathrm{u}$ for PO is limited, with only two distinct values (8.4\,K and 15.7\,K), we also fit synthetic spectra to each individual hyperfine line to find the parameters derived from the rotation diagram fit agree well with the data (see Figure\,\ref{fig:PO}). Similarly, using the calculated $T_\mathrm{ex}$ value of 4.6\,K for PO, we fit synthetic spectra to the PO$^{+}$ transitions and find a column density roughly two orders of magnitude lower than for PO, at $N_\mathrm{tot} = 3.02^{+0.15}_{-0.12} \times 10^{10}$\,cm$^{-2}$. All column densities are reported in Table\,\ref{table:3}.

As seen before in the literature (e.g., \citealt{2016ApJ...822L..30F, 2022ApJ...934..153W}) this emission is expected to be subthermally excited.
Future work to model these P-bearing species in the optically thick limit and with non-LTE codes in conjunction with a better understanding of the physical structure (e.g., volume density) of the emitting gas can shed light on this, but is beyond the scope for this current Letter. 

Additionally, we note that it is still unclear if the PN, PO and PO$^{+}$ emission comes from a compact region around the core or if they trace the larger-scale SiO emission. Therefore, we assume the emission fills our beams ($f = 1$) in all cases and thus our $N$ values are likely lower limits. We do note that for PN if we assume a smaller filling fraction (using the same source size of 51.5\,arcsec as found for methanol in \citealt{2024MNRAS.533.4104S}), the $T_\mathrm{ex}$ drops to 6.1\,K and the total column density is slightly raised to $N_\mathrm{tot} = 1.2 \times 10^{12}$\,cm$^{-2}$.

\begin{figure}
\begin{center}$
\centering
\begin{array}{c}
\includegraphics[width=80mm]{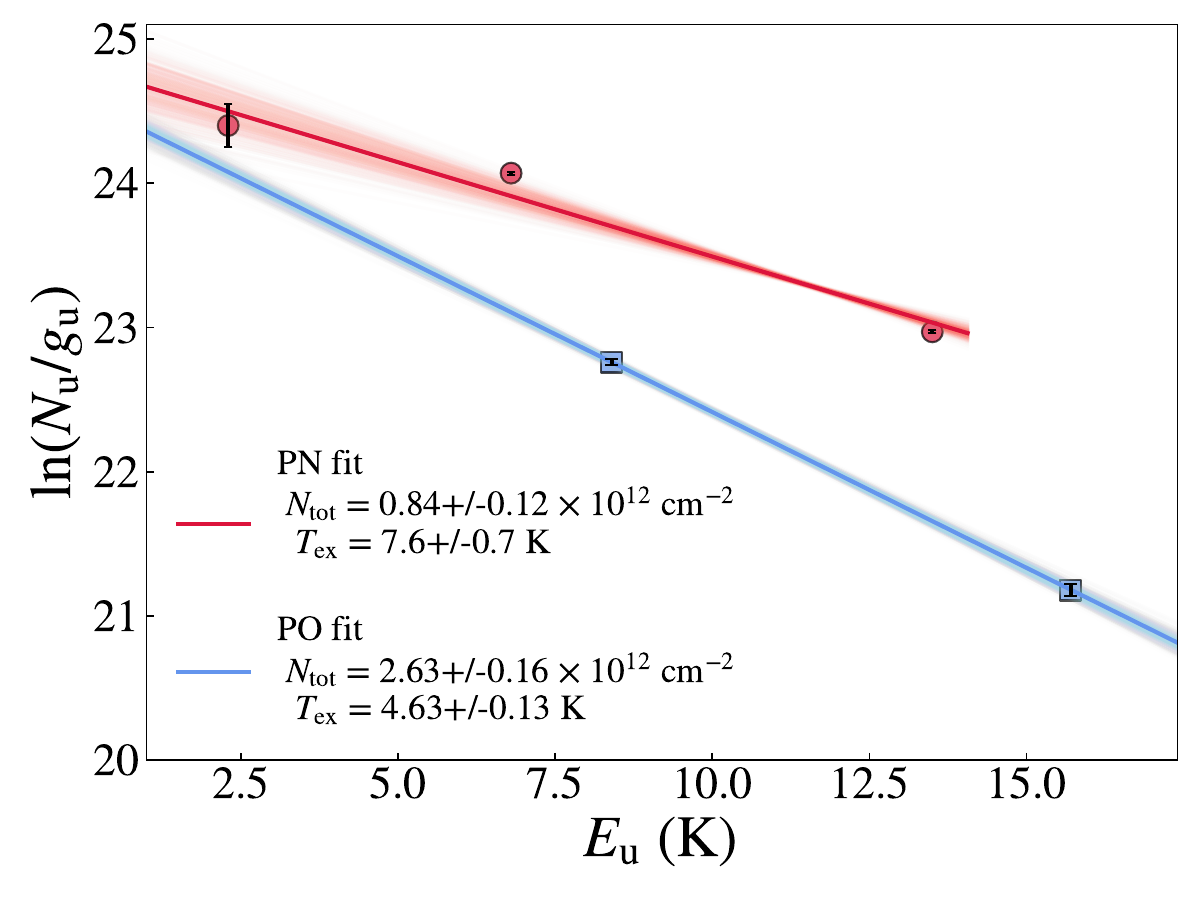} 
\end{array}$
\end{center}
    \caption{Rotation diagram for PN (red circle points) for the three observed $J$ transitions as well as PO (blue square points) for the two observed $J$ transitions.  From the Monte Carlo simulations, the best-fit is shown as the solid line and the samples are shown as faded lines. The calculated column densities ($N_\mathrm{tot}$) and excitation temperatures ($T_\mathrm{ex}$) are in the bottom left corner. } \label{fig:rot}
\end{figure}


\begin{table}
\caption{\label{table:2} Prestellar and starless cores in Perseus with PN(1-0) and SiO(1-0) constraints from the Yebes 40\,m.}
\setlength{\tabcolsep}{2pt}
\begin{center}
\begin{tabular}{lccclc}
\hline\hline
 & & & \multicolumn{2}{c}{PN(1-0)} & \multicolumn{1}{c}{SiO(1-0)}  \\ 
Core\,\# & RA&Dec & Detect? & $N$ $\times10^{12}$  & Detect? \\
  &  (J2000)& (J2000) &   &  [cm$^{-2}$] &  \\
\hline
264 & 03:28:47.14 & +31:15:11.4&  \textit{\sffamily x} & $<$0.36   & \checkmark$^*$  \\
 317 &  03:29:04.93 & +31:18:44.4& \textit{\sffamily x} & $<$0.33   & \checkmark$^*$  \\
 321 & 03:29:07.17 & +31:17:22.1 & \textit{\sffamily x} & $<$0.33   & \checkmark$^*$  \\
 326 & 03:29:08.97 &  +31:15:17.2 & \checkmark & 0.84$\pm0.12$ & \checkmark  \\
 413 & 03:30:46.74 &	+30:52:44.8 & \textit{\sffamily x} & $<$0.39 &  \textit{\sffamily x}  \\
 504  & 03:33:25.31 &	+31:05:37.5 & \textit{\sffamily x} & $<$0.38 &  \textit{\sffamily x} \\
 615 & 03:40:14.92 & +32:01:40.8&  \textit{\sffamily x} & $<$0.41   & \textit{\sffamily x} \\ 
 627 &03:40:49.53 & +31:48:40.5 & \textit{\sffamily x} & $<$0.40  & \textit{\sffamily x}  \\ 
 709 & 03:43:38.06 & +32:03:07.4 &\textit{\sffamily x} & $<$0.31  & \checkmark{$^*$} \\ 
 715 & 03:43:46.34 & +32:01:43.5 &\textit{\sffamily x} & $<$0.38 & \checkmark{$^*$} \\ 
 746 & 03:44:14.38 & +31:58:00.7 &\textit{\sffamily x} & $<$0.38   & \textit{\sffamily x} \\ 
 752 & 03:44:23.10 & +32:10:01.0  & \textit{\sffamily x} & $<$0.41  & \checkmark{$^*$} \\
 768 & 03:44:48.83 & +32:00:31.6 & \textit{\sffamily x} & $<$0.28  & \textit{\sffamily x} \\
799 &  03:47:31.31 & +32:50:56.9&\textit{\sffamily x} & $<$0.48 &   \textit{\sffamily x} \\
 800 &  03:47:38.97 & +32:52:16.6& \textit{\sffamily x} & $<$0.39 &  \textit{\sffamily x} \\
\hline

\end{tabular}
\end{center}
\tablecomments{ The variation in 3$\sigma$ upper limits reflect the varying noise levels from source to source. Not shown is non-detection (--) of PO$^{+}$(1-0) in each source. $^*$Cores with a SiO(1-0) line brightness $T_\mathrm{mb} <$ 0.3\,K (note that for Pers326 $T_\mathrm{mb}> 1$\,K; see Figure\,\ref{fig:yebesSiO}). } 
\end{table}

\section{Results and Discussion} \label{sec:discuss}

We report the first detection of P-bearing molecules toward a shocked low-mass starless core. These observations provide new insight into the interstellar phosphorus budget in the earliest stage of star formation.

As already mentioned, we find kinematic correlation between the shock tracer SiO to the PN, PO and PO$^{+}$ emission in starless core Pers326 (Figures\,\ref{fig:yebesSiO},\,\ref{fig:yebesPN},\,\ref{fig:PO}\, and \ref{fig:POplus}). Previous observations from high-angular resolution mapping studies both in high-mass \citep{2020MNRAS.492.1180R, 2024A&A...682A..74F} and low-mass \citep{2022ApJ...927....7B} protostellar objects reveal PN and PO are also typically spatially correlated with SiO. Within the NGC\,1333 region (Figure\,\ref{fig:outflow_info}), the widespread narrow-line SiO (mapped by \citealt{1998ApJ...504L.109L}) is hypothesized to come either from the sputtering of dust grains at the surface of the shocked gas fragments, or from the deflected outflow material that propagates in the lower density gas. 

There is still no consensus, however, on whether this shocked gas does indeed originate from nearby protostellar outflows, or rather from large-scale cloud collisions that have produced `trains' of shocks \citep{2022MNRAS.512.5214D}. Despite there being no direct evidence of these shock trains in the northern region of NGC\,1333 (where SVS13 A and Pers326 are located), we do note that the proposed shock front of the northern bubble hypothesized by \cite{2022MNRAS.512.5214D} coincides with the location of Pers326, which is directly along the knots of bright SiO emission mapped by \cite{1998ApJ...504L.109L} (see Figure 7 in \citealt{2022MNRAS.512.5214D}). Regardless of the precise origin, this correlation of P-bearing molecules to SiO toward Pers326 supports the idea that shock and/or outflow interactions are needed to get phosphorus off dust grains.

Out of the 15 total starless and prestellar cores surveyed with the Yebes 40 m, Pers326 remains the only source with strong ($T_\mathrm{mb}$ $>$ 1\,K) SiO(1-0) emission (Figure\,\ref{fig:yebesSiO} and Table\,\ref{table:2}). It is also the only core directly along the outflow direction of a nearby protostar, SVS13 A, in NGC\,1333 (see Figure\,\ref{fig:outflow_info}; \citealt{2013ApJ...774...22P}). An outlier compared to the other starless and prestellar cores in the sample, even those with low-level ($T_\mathrm{mb}<0.3$\,K) SiO(1-0) emission, Pers326 likely has experienced a higher degree of energetic processes that, first, can get more SiO off the grains (e.g., via sputtering; \citealt{1997A&A...322..296C}) and, second, produce enough PN to be detectable at our telescope sensitivity ($N >0.28 - 0.58 \times 10^{12}$\,cm$^{-2}$; Table\,\ref{table:2}). And, while P-bearing species have also not been detected in any other low-mass starless or prestellar core in the literature (e.g., L1544; \citealt{Furuya_2024}), the presence of PN, PO and PO$^{+}$ even in one source, Pers326, suggests that phosphorus is locked up in the solids at this early stage, ready to be released into the gas-phase.  

The clear detection of bright ($T_\mathrm{mb} >30$\,mK) PN and PO lines in a region not associated with any protostellar object also suggests the emission is likely more extended than what has been previously observed as compact spots associated with protostellar outflows. 
Since PN is also observed in more massive quiescent clouds \citep{2016ApJ...822L..30F}, it is likely that the gas we observe in Pers326 first originated in shocked material from the nearby protostellar ouflows (i.e., from SVS13\,A, IRAS\,4A and IRAS\,2A in Figure\,\ref{fig:outflow_info}; \citealt{1998ApJ...504L.109L, 2000A&A...361..671K, 2013ApJ...774...22P}) and has remained in the gas-phase even at low velocities and low temperatures. While it may also be possible that this shocked material was created from an expanding bubble (likely an old supernova remnant) colliding with the molecular cloud, the source of this bubble is more uncertain \citep{2022MNRAS.512.5214D}.
The spatial distribution of SiO with these P-bearing molecules should be studied toward Pers326 in order to say more. 

We do find a `global' abundance ratio of PO/PN of $3.1^{+0.4}_{-0.6}$, which lies at the high end of the range of values reported by \cite{2022ApJ...934..153W} for their sample of protostars, with ratios from 0.6 to 2.2. Our ratio is also consistent with the PO/PN ratios of $\sim3$, 1.8 and 3, measured in the low-mass outflow L1157-B1 \citep{2016MNRAS.462.3937L} and the high-mass star-forming regions W51 and W3(OH) \citep{2016ApJ...826..161R}, respectively. Therefore, in this early phase of star formation, the phosphorus chemistry likely proceeds similarly as in the protostellar environment. 

The formation of PO and PN starts from parent phosphorus carriers, proposed by \cite{2018ApJ...862..128J} to be phosphine (PH$_3$) on ice mantles, though this species has not yet been detected in any star-forming environment\,(e.g., \citealt{Furuya_2024}). Perhaps more likely, phosphate minerals or phosphorus oxides could be main grain carriers \citep{2021ApJ...906...55B, 2022ApJ...927....7B}. Phosphorus could also be found in atomic form, as is the case for cometary ices (e.g., \citealt{2016SciA....2E0285A}). Regardless, once P is available to form into molecules the gas-phase, there have been several proposed ways to get the abundance of PO higher than PN. At high shock speeds ($v_\mathrm{s}\geq$\,40 km/s) PO can be efficiently formed from either from PN + O \citep{2018ApJ...862..128J} or P + OH \citep{2021ApJ...922..169G}. More relevant for our quiescent source Pers326, it was recently found by \cite{2024ApJ...963..142G} that at lower shock velocities the most efficient formation pathway for PO is P + O$_2$. Still, shock models typically produce PO/PN ratios at less than unity (e.g., \citealt{2018ApJ...862..128J, 2021AJ....162..119S}), and \cite{2023ApJ...956...47F} argue this is due to inaccuracies in reaction rate coefficients and a lack of important destruction routes for PN. 

An additional challenge to our understanding of phosphorus chemistry in shocked regions is the first detection of the ion PO$^{+}$ in a low-mass star-forming environment, which we present here (Figure\,\ref{fig:POplus}). The first and only previous detection of PO$^{+}$ was made toward the massive dark molecular cloud G+0.693 in the Galactic Center, where a PO$^+$/PO ratio of $0.12\pm0.03$ was found \citep{2022FrASS...9.9288R}. To explain an enhanced abundance of PO$^+$, the authors found models with high cosmic ionization rates of $\zeta = 10^{-15}-10^{-14}\, \mathrm{s}^{-1}$ were better able to match the observations. We find for Pers326, a PO$^{+}$/PO ratio ten times lower, at $0.0115^{+0.0008}_{-0.0009}$ (see Table\,\ref{table:3}), which suggests PO$^{+}$ production can still be efficient at lower cosmic ionization rates, which have been calculated to be $\zeta \sim 10^{-17}-10^{-16}\, \mathrm{s}^{-1}$ toward Pers326 (see NGC\,1333 map of $\zeta$ in \citealt{2024A&A...686A.162P}).

The `trail' of phosphorus in low-mass star formation extends now from the earliest prestellar stage, to the protostellar stage, as well as to local solar system objects, such as comets and asteroids. Recent samples from the asteroid Ryugu reveal the presence of phosphorus-rich grains formed from an increased availability of P--O-rich ions, which may have offered a pathway toward the formation of organophosphorus compounds, including organophosphates known to be part of biomolecules such as RNA, DNA and ATP \citep{2024NatAs...8.1529P}. 

The comet 67P/Churyumov-Gerasimenko, or 67/P, provides a more direct quantitative comparison of primordial phosphorus abundance with respect to the `mother' grain-produced organic molecule methanol, where a PO/CH$_3$OH ratio is found to be $\sim 5\%$ \citep{Rubin_2019}. In the starless core Pers326 we find a (PN+PO)/CH$_3$OH ratio of $0.69 ^{+0.07}_{-0.08} \%$, which falls toward the low-end of values for the sample of protostars from \cite{2022ApJ...934..153W} at 0.7 to 2.7\,$\%$. This rough indicator of `total' P abundance suggests that we see a somewhat lower fraction of volatile phosphorus in Pers326 compared to protostars, but in general is consistent with the majority of P being locked up in solids. 
Larger sample sizes are needed to better assess whether or not the previously shocked quiescent gas in the earliest phase of low-mass star formation is directly related to the phosphorus budget observed in comets and asteroids. 

\begin{table}
\caption{\label{table:3} Column densities and abundance ratios for Pers326.}
\begin{center}
\setlength{\tabcolsep}{5pt}
\begin{tabular}{lccc}
\hline\hline
  & $N$ [cm$^{-2}$] &  $X[\mathrm{H_2}]$  \\
\hline
PN &  $0.84 \pm 0.12\times10^{12}$ & $1.06^{+0.32}_{-0.23}\times10^{-11}$ \\
PO &  $2.63 \pm 0.16\times10^{12}$ &  $3.3^{+0.6}_{-0.9}\times10^{-11}$ \\
PO$^{+}$ & $3.02 ^{+0.15}_{-0.12} \times10^{10}$ &  $3.8^{+0.7}_{-1.0}\times10^{-13}$  \\  
\hline 
\hline 
PO/PN & $3.1^{+0.4}_{-0.6}$ & -- \\
PO$^{+}$/ PO & $0.0115^{+0.0008}_{-0.0009}$ & --\\
(PN+PO)/CH$_3$OH &  $0.0069 ^{+0.0007}_{-0.0008} $ & --\\ 
\hline
\end{tabular}
\end{center}
\tablecomments{Values for a beam-averaged H$_2$ column density ($7.9 \pm 1.6 \times10^{22}$ cm$^{-2}$) and CH$_3$OH column density ($5.03 ^{+0.52}_{-0.43} \times10^{14}$ cm$^{-2}$) are from \citealt{2024MNRAS.533.4104S}. } 
\end{table}

\section{Conclusions} \label{sec:conclude}

An unbiased (e.g., large bandwidth) line survey with the Yebes 40\,m of more than a dozen starless and prestellar cores in the Perseus Molecular Cloud \citep{2024MNRAS.533.4104S} led to the serendipitous detection of the phosphorus molecule PN in the Pers326 starless core. Follow-up observations with the IRAM 30\,m confirmed the PN detection and also detected multiple transitions of the chemically related PO molecule as well as a single transition of the ion PO$^{+}$. These observations are the first of their kind toward any shocked low-mass starless core. 

The presence of PN is kinematically correlated with bright ($T_\mathrm{mb} >1$\,K) SiO(1-0) emission and is not detected in any other source from the 15-core sample, even those with low-level ($T_\mathrm{mb} <0.3$\,K) SiO(1-0) emission. Due to this correlation, we infer shocks are necessary for the detection of gas-phase P toward starless cores. Still, the narrow (FWHM $\sim 0.4-1.0$\,km/s) PN and PO emission lines toward Pers326, which contains no central protostar, suggest that the phosphorus emission has originated from gas that was previously shocked and/or influenced by shocks, not the shock material itself. 

The column densities of PN, PO and PO$^{+}$ were also constrained in the optically thin limit using LTE radiative transfer methods. The PO/PN ratio is $3.1^{+0.4}_{-0.6}$, consistent with other literature values. The PO$^{+}$/PO ratio is $0.0115^{+0.0008}_{-0.0009}$, a factor of ten lower than previously found toward a Galactic Center molecular cloud \citep{2022FrASS...9.9288R}, suggesting high ($>10^{-15}$ s$^{-1}$) cosmic ionization rates are not exclusively needed to efficiently form PO$^{+}$. Updated chemical models that can account for enhanced PO/PN ratios as well as the inclusion of PO$^{+}$ pathways are needed. 

We are now able to probe phosphorus across a wider range of stages in low-mass star-formation, from cold starless cores to comets. 
The abundances of PN and PO show similarities throughout the stages of low-mass star formation when normalized to CH$_3$OH, suggesting the phosphorus inventory may be set early. To better trace this precursor prebiotic chemistry, increased detection statistics and spatially resolved observations of P-bearing species in and around low-mass shocked starless cores are needed.  

\begin{acknowledgments}
We thank the anonymous reviewer for their informative comments and feedback. The authors also thank Anthony Remijan and V\'ictor Rivilla for helpful discussion and suggestions. 
S.S. acknowledges the National Radio Astronomy Observatory is a facility of the National Science Foundation operated under cooperative agreement by Associated Universities, Inc. 
I.J-.S and A.M. acknowledge funding from grant PID2022-136814NB-I00 funded by the Spanish Ministry of Science, Innovation and Universities/State Agency of Research MICIU/AEI/ 10.13039/501100011033 and by “ERDF/EU”, and from the ERC grant OPENS (project number 101125858) funded by the European Union.
Y.S. was supported by National Science foundation Astronomy and Astrophysics grant AST-2205474. R.T.G. thanks the National Science Foundation for funding through the Astronomy \& Astrophysics program (grant number 2206516).  

Observations carried out with the Yebes 40\,m telescope (22A022 and 23A025). Paula Tarr\'io and Alba Vidal Garc\'ia carried out the observations and the first inspection of the data quality. The 40\,m radio telescope at Yebes Observatory is operated by the Spanish Geographic Institute (IGN; Ministerio de Transportes, Movilidad y Agenda Urbana). Yebes Observatory thanks the European Union’s Horizon 2020 research and innovation programme for funding support to ORP project under grant agreement No 101004719. This work is also based on observations carried out under project number 024-24 with the IRAM 30\,m telescope. IRAM is supported by INSU/CNRS (France), MPG (Germany) and IGN (Spain).

\end{acknowledgments}

\software{APLpy \citep{2012ascl.soft08017R}, astropy \citep{2013A&A...558A..33A,2018AJ....156..123A}, GILDAS \citep{2005sf2a.conf..721P, 2013ascl.soft05010G}, GILDAS-CLASS Pipeline \citep{2023MNRAS.519.1601M}, Matplotlib \citep{2007CSE.....9...90H}, NumPy \citep{2020arXiv200610256H}, \href{https://github.com/andresmegias/richvalues}{richvalues}, SciPy \citep{2020SciPy_NMeth}, Pyspeckit \citep{2011ascl.soft09001G, 2022AJ....163..291G}.  }







\bibliography{references.bib}{}
\bibliographystyle{aasjournal}



\end{document}